\documentclass[11pt]{article}
\usepackage[colorlinks=true,citecolor=blue,linkcolor=blue]{hyperref}
\usepackage{multirow}
\usepackage[left=2cm,right=2cm,top=2cm,bottom=2cm]{geometry}
\usepackage{cite}
\usepackage{psfrag}
\usepackage[demo]{graphicx}
\usepackage{graphicx}
\usepackage{graphics}
\usepackage{epsfig}
\usepackage{booktabs}
\usepackage{amsmath,amsfonts,amssymb}
\usepackage{pstcol,pst-fill,pst-grad}
\usepackage{pstricks,pst-fill,pst-grad}
\usepackage{euscript}
\usepackage{pstricks}
\usepackage{wrapfig}
\usepackage{slashed}
\usepackage[toc,page]{appendix}
\usepackage{here}

\begin{document}
	\title{\vspace{-3cm}
		\hfill\parbox{4cm}{\normalsize \emph{}}\\
		\vspace{1cm}
		{Laser-induced proton decay}}
	\vspace{2cm}
	
	\author{M. Ouhammou,$^1$ M. Ouali,$^1$ S. Taj,$^1$ R. Benbrik,$^2$ and B. Manaut$^{1,}$\thanks{Corresponding author, E-mail: b.manaut@usms.ma} \\
		{\it {\small$^1$ Polydisciplinary Faculty, Laboratory of Research in Physics and Engineering Sciences,}}\\
		{\it {\small Team of Modern and Applied Physics, Sultan Moulay Slimane University,}}\\
		{\it {\small Beni Mellal, 23000, Morocco.}}\\	
		{\it {\small$^2$ LPFAS, Polydisciplinary Faculty of Safi, UCAM, Morocco}}\\				
	}
	\maketitle \setcounter{page}{1}
\date{}
\begin{abstract}
In this research paper, we investigate the decay of the proton into neutron, positron and electron neutrino in the presence of an external electromagnetic field with circular polarization. Different physical quantities related to this decay process, such as proton's decay rate and its lifetime, are calculated based on the S-matrix approach. The proton and positron are treated as Dirac-Volkov states, while the neutron and electron neutrino are free-states. 
We have found that, though it can not occur in vacuum, the proton's decay process into neutron, positron and electron neutrino becomes possible in the presence of laser field with high intensities near or close to the Schwinger limit. In addition, near this limit and for some frequencies, the proton's lifetime can be comparable to that of the neutron, and the required laser strength, from which this decay becomes possible, depends on the chosen laser source.
\end{abstract}
Keywords: Electroweak interaction, proton's decay, laser-induced processes.
\section{Introduction}
The laser, as a source of pure energy in the form of monochromatic and coherent photons, has become an irreplaceable tool in many area such as medical technology, metallurgy and the electronics industry. 
Since its invention by Theodore Maiman in 1960 \cite{Maiman}, this extraordinary light source has had a breath-taking career, both in science and commercial applications. The various applications of the laser were consequences of its natural evolution led by both experimental and theoretical studies. The field of laser-matter interactions is receiving increasing attention in enormous fields of physics such as optics, atomic physics, nuclear physics, quantum electrodynamics, and recently in high energy physics \cite{Piazza}. In general, these electromagnetic interactions with matter may be classified into two categories. 
The first category is known as laser-assisted interactions \cite{as1,as2,as3,as5,as6,as7,as8}, where the latter can occur also in the absence of the laser field. 
It is found that the laser field may affect not only the cross section of the interaction process but also the decay width and lifetime of a particle. For instance, in \cite{Pion} the authors showed that the lifetime of the pion increases and its decay width decreases by the presence of a circularly polarized laser field. In addition, similar results are found for charged kaon and gauge bosons decay \cite{had-decay, lept-decay,W-decay}.
However, in the second category (known as laser-induced processes), the interaction is induced by the electromagnetic field and can not occur without it \cite{Ind1,Ind2,Ind3,Ind4,Ind5,Ind6,Ind7,Ind8}. In this case, the lifetime of a particle can be modified under the influence of acceleration by an external field, and new decay and scattering processes may become possible.

The fact that the entirety of our universe appears to be made of matter and not antimatter is a good reason to expect that the proton may decay. However, for many years, the Kamioka Nucleon Decay Experiment \cite{Kamiokande} ran looking for proton decay, and it yields no evidence for it. In addition, in \cite{search for proton decay}, the authors conducted a research about proton decay in the Super-Kamiokande experiment detector via four modes ($p\rightarrow e^{+}\pi^{0}$, $p\rightarrow \mu^{+}\pi^{0}$, $p\rightarrow \bar{\nu}K^{+}$ and $p\rightarrow \bar{\nu}K^{*}$), using a 141.3-kiloton year exposure of the Super-Kamiokande water Cherenkov detector, and they did not find any hints for proton decay via any of these modes. However, these experiments have been able to establish lower bounds on the the proton half-lifetime. Recently, a precise results come from the Super-Kamiokande water Cherenkov radiation detector expected the lifetime of the proton as $2.4\times 10^{34}$ and $1.6\times 10^{34}$ years via positron and antimuon decays, respectively \cite{lifetime}. Consequently, the proton is regarded as a stable particle in the standard model.

Theoretically and according to conservation laws of particle physics, the proton can only decay, though its high lifetime, into lighter subatomic particles than itself such as a neutral pion and a positron, and it cannot decay into a neutron or any other combination of three quarks. In \cite{Keitel}, the authors studied the probability of inducing the proton's decay into neutron, positron and electron neutrino by a linearly polarized laser wave, and they found that this process can occur when the proton experiences a laser field strength about ten times the Schwinger limit.
However, non linear effects of quantum electrodynamics become more important beyond this limit \cite{Ritus}.
Against this background, we study in this paper the probability of reducing the proton's lifetime and turning it into an unstable particle that decays into heavier particles such as the neutron in association with a positron and electron neutrino by exposing the proton into a powerful laser pulse with circular polarization. 

The rest of this paper is organized as follows: The next section deals with the theoretical calculation of the proton decay rate and its lifetime inside an electromagnetic wave with circular polarization. Then, we analyze and discuss the obtained numerical results about the decay rate, lifetime and the effective mass of the proton as a function of the laser parameters. A short conclusion is given in the last section. We use natural units throughout this paper such that $\hbar=c=1$. The choice made for Livi-Civita tensor is that $\epsilon^{0123}=1$, and the metric $g^{\mu\nu}$ is chosen as $g^{\mu\nu}=(1,-1,-1,-1)$.
\section{THEORETICAL CALCULATION}
In this part, we consider the theoretical analysis of the decay rate and lifetime of the proton decay into a neutron, a positron, and an electron-neutrino in the presence of a laser field. This process can be described as follows:
\begin{equation}
P(q_{1})\longrightarrow N(p_{2})+e^{+}(q_{3})+\nu_{e}(p_{4}),
\label{eq1}
\end{equation}
where the arguments label the associated momenta.
The proton is considered as spin-$\frac{1}{2}$ particle.
In the presence of an electromagnetic potential the wave functions of the relativistic proton and positron can be derived by solving the following Dirac equation:
\begin{equation}
(i\slashed{\partial}-e\slashed{A}-m)\psi_{e}(x)=0.
\label{eq2}
\end{equation}
For circular polarization, $A^{\mu}(\phi)$ has the following expression:
\begin{equation}
A^{\mu}(\phi)=a_{1}^{\mu}\cos(\phi)+a_{2}^{\mu}\sin(\phi)\qquad;\qquad \phi=(k.x),
\label{eq3}
\end{equation}
where the polarization 4-vectors $ a_{1}^{\mu}=|\mathbf{a}|(0,1,0,0) $ and $ a_{2}^{\mu}=|\mathbf{a}|(0,0,1,0) $ verify the following conditions: $ (a_{1}.a_{2})=0 $ and $ a_{1}^{2}=a_{2}^{2}=a^{2}=|\mathbf{a}|^{2}=(\varepsilon_{0}/\omega)^{2} $, with $ \varepsilon_{0} $ is the amplitude of the electric field. $ k=(\omega,\mathbf{k}) $ is the electromagnetic wave 4-vector such that $ (k^{2}=0) $ and $ k_{\mu}A^{\mu}=0 $ (Lorentz gauge condition), and it is chosen to be along the $z$-axis. $ \phi $ is the phase of the laser field and $ \omega $ its frequency.
The lowest-order scattering matrix element for the laser-induced proton decay reads \cite{Greiner}:
\begin{equation}
S_{fi}=\dfrac{-iG_{F}}{\sqrt{2}}\int d^{4}x \big[\bar{\Psi}_{p_{2},s_{2}}(x)\gamma^{\mu}(g_{v}-g_{a}\gamma^{5})\Psi_{p_{1},s_{1}}(x)\big]\big[\bar{\Psi}_{p_{4},s_{4}}(x)\gamma_{\mu}(1-\gamma^{5})\Psi_{p_{3},s_{3}}(x)\big],
\label{eq4}
\end{equation}
where $ G_{F}=1.16637\pm 0.00002\times 10^{-5} GeV^{-2} $ is the Fermi coupling constant. $ g_{v} $ and $ g_{a} $ are, respectively, the vector and axial-vector coupling constants. The incoming proton is positively charged. Therefore, its wave function is given by the relativistic Dirac-Volkov function normalized to the volume V as follows \cite{Volkov}:
\begin{equation}
\psi_{p_{1},s_{1}}(x)= \Big[1+\dfrac{e \slashed k \slashed A}{2(k.p_{1})}\Big] \frac{u(p_{1},s_{1})}{\sqrt{2Q_{1}V}} \exp^{iS(q_{1},s_{1})},
\label{eq5}
\end{equation}
where
\begin{equation}
S(q_{1},s_{1})=- q_{1}x -\frac{e(a_{1}.p_{1})}{(k.p_{1})}\sin\phi + \frac{e(a_{2}.p_{1})}{(k.p_{1})}\cos\phi .
\label{eq6}
\end{equation}
$u(p_{1},s_{1})$ is the Dirac bispinor for the free charged proton where the free momentum $ p_{1} $ and spin $ s_{1} $ are satisfying the following relation $ \sum_{s_{1}}u(p_{1},s_{1})\bar{u}(p_{1},s_{1})=\slashed p_{1}+m_{p_{1}} $, with $ m_{p_{1}} $ is the rest mass of the charged proton. The 4-vector $ q_{1}=p_{1}+e^{2}a^{2}/2(k.p_{1})k $ is the quasi-momentum that the charged proton acquire in the presence of the electromagnetic field.
Similarly, the Dirac-Volkov state of the positron in the laser field is expressed as follows \cite{Volkov}:
\begin{equation}
\psi_{p_{3},s_{3}}(x)= \Big[1+\dfrac{e \slashed k \slashed A}{2(k.p_{3})}\Big] \frac{v(p_{3},s_{3})}{\sqrt{2Q_{3}V}} \exp^{iS(q_{3},s_{3})},
\label{eq7}
\end{equation}
where
\begin{equation}
S(q_{3},s_{3})=+ q_{3}x +\frac{e(a_{1}.p_{3})}{(k.p_{3})}\sin\phi - \frac{e(a_{2}.p_{3})}{(k.p_{3})}\cos\phi ,
\label{eq8}
\end{equation}
with $ q_{3}=(Q_{3},\mathbf{q_{3}}) $ is the effective four-momentum of the positron, and it is related to its corresponding free momentum by the following equation:
\begin{equation}
q_{3}=p_{3}+\dfrac{e^{2}a^{2}}{2(k.p_{3})}k .
\label{eq9}
\end{equation}
 The outgoing neutron and electron-neutrino are electrically neutral. Consequently, they do not interact with the laser field, and they are described by free-states as follows:
\begin{equation}
\begin{cases}
\psi_{p_{2},s_{2}}(x)=\dfrac{1}{\sqrt{2E_{2}V}}u(p_{2},s_{2})e^{-ip_{2}.x}&\\
\psi_{p_{4},s_{4}}(x)=\dfrac{1}{\sqrt{2E_{4}V}}u(p_{4},s_{4})e^{-ip_{4}.x} ,
\end{cases}
\label{eq10}
\end{equation}
where $ p_{2} $ and $ p_{4} $ indicate the momentum of the neutron and electron-neutrino, respectively. $ s_{i}(i=2,4) $ and $ E_{i}(i=2,4) $ denote their spins and energies, respectively.
Inserting the equations (\ref{eq5}), (\ref{eq7}) and (\ref{eq10}) into the equation (\ref{eq4}) and after some algebraic calculations, we find that the S-matrix element can be written as:
\begin{equation}
S_{fi}=\dfrac{-iG_{F}}{\sqrt{2}}\dfrac{1}{4V^{2}\sqrt{Q_{1}Q_{3}E_{2}E_{4}}}\sum_{r=-\infty}^{+\infty}M_{fi}^{r}(2\pi)^{4}\delta^{4}(p_{2}+p_{4}+q_{3}-q_{1}-rk),
\label{eq11}
\end{equation}
where $ r $ is the number of exchanged photons. The quantity $ M_{fi}^{r} $ is defined as:
\begin{eqnarray}
M_{fi}^{r}&=&\nonumber\Big[\bar{u}(p_{2},s_{2})\big[C_{0}B_{0r}(z)+C_{1}B_{1r}(z)+C_{2}B_{2r}(z)\big]u(p_{1},s_{1})\Big]\\&\times &\Big[\bar{u}(p_{4},s_{4})\big[D_{0}B_{0r}(z)+D_{1}B_{1r}(z)+D_{2}B_{2r}(z)\big]u(p_{3},s_{3})\Big],
\label{eq12}
\end{eqnarray}
with:
\begin{equation}
\begin{cases}
C_{0}=\gamma^{\mu}(g_{v}-g_{a}\gamma^{5})&\\
C_{1}=\dfrac{e}{2(kp_{1})}\gamma^{\mu}(g_{v}-g_{a}\gamma^{5})\slashed k\slashed a_{1}&\\
C_{2}=\dfrac{e}{2(kp_{1})}\gamma^{\mu}(g_{v}-g_{a}\gamma^{5})\slashed k\slashed a_{2}
\end{cases}
 \text{and} \qquad
 \begin{cases}
D_{0}=\gamma^{\mu}(1-\gamma^{5})&\\
D_{1}=\dfrac{e}{2(kp_{3})}\gamma^{\mu}(1-\gamma^{5})\slashed k\slashed a_{1}&\\
D_{2}=\dfrac{e}{2(kp_{3})}\gamma^{\mu}(1-\gamma^{5})\slashed k\slashed a_{2}
\end{cases}.
\label{eq13}
\end{equation}
The coefficients $ B_{0r}(z) $, $ B_{1r} (z)$ and $ B_{2r} (z)$ are explicitly expressed in terms of Bessel functions as follows:
\begin{equation}
\left.
  \begin{cases}
     B_{0r}(z) \\
      B_{1r}(z) \\
      B_{2r}(z)
  \end{cases}
  \right\} = \left.
  \begin{cases}
     J_{\frac{r}{2}}(z)e^{-i\frac{r}{2}\phi _{0}}\\
    \frac{1}{2}\big(J_{\frac{r}{2}+1}(z)e^{-i(\frac{r}{2}+1)\phi _{0}}+J_{\frac{r}{2}-1}(z)e^{-i(\frac{r}{2}-1)\phi _{0}}\big) \\
     \frac{1}{2\, i}\big(J_{\frac{r}{2}+1}(z)e^{-i(\frac{r}{2}+1)\phi _{0}}-J_{\frac{r}{2}-1}(z)e^{-i(\frac{r}{2}-1)\phi _{0}}\big)
  \end{cases}
  \right\} .
  \label{eq14}
\end{equation}
The argument of the Bessel function and its phase are expressed by:
$z=\sqrt{(\dfrac{\alpha_{1}}{2})^{2}+(\dfrac{\alpha_{2}}{2})^{2}}$ and $\phi_{0}= \arctan(\frac{\alpha_{2}}{\alpha_{1}})$, where:
\begin{center}
$\alpha_{1}=\Big(\dfrac{e(a_{1}.p_{1})}{(k.p_{1})}-\dfrac{e(a_{1}.p_{3})}{(k.p_{3})}\Big)$ \qquad ;\qquad $\alpha_{2}=\Big(\dfrac{e(a_{2}.p_{1})}{(k.p_{1})}-\dfrac{e(a_{2}.p_{3})}{(k.p_{3})}\Big)$.
\end{center}
The decay rate of the proton per particle and per time into the final states is obtained by: Squaring the scattering-matrix element given by equation (\ref{eq11}), summing over the polarization of the final states, averaging over the initial one, and finally dividing by the time T. It is mathematically expressed as:
\begin{equation}
d\Gamma=\dfrac{1}{T}|S_{fi}|^{2}V\int\dfrac{d^{3}q_{3}}{(2\pi)^{3}}V\int\dfrac{d^{3}p_{2}}{(2\pi)^{3}}V\int\dfrac{d^{3}p_{4}}{(2\pi)^{3}} ,
\label{eq15}
\end{equation}
where
\begin{eqnarray}
|S_{fi}|^{2}&=&\nonumber\dfrac{G_{F}^{2}}{2}\dfrac{1}{16V^{4}Q_{1}Q_{3}E_{2}E_{4}}\sum_{r=-\infty}^{+\infty}Tr\Big[(\slashed p_{2}+m_{N})\Delta^{r}(\slashed p_{1}+m_{P})\bar{\Delta^{r}}\Big]\\&\times & Tr\Big[(\slashed p_{4}+m_{\nu})\Lambda^{r}(\slashed p_{3}-m_{e})\bar{\Lambda^{r}}\Big](2\pi)^{4}VT\delta^{4}(p_{2}+p_{4}+q_{3}-q_{1}-rk) ,
\label{eq16}
\end{eqnarray}
with 
\begin{equation}
\begin{cases}
\Delta^{r}=C_{0}B_{0r}(z)+C_{1}B_{1r}(z)+C_{2}B_{2r}(z)&\\
\bar{\Delta^{r}}=\bar{C_{0}}B^{*}_{0r}(z)+\bar{C_{1}}B^{*}_{1r}(z)+\bar{C_{2}}B^{*}_{2r}(z)&\\
\Lambda^{r}=D_{0}B_{0r}(z)+D_{1}B_{1r}(z)+D_{2}B_{2r}(z)&\\
\bar{\Lambda^{r}}=\bar{D_{0}}B^{*}_{0r}(z)+\bar{D_{1}}B^{*}_{1r}(z)+\bar{D_{2}}B^{*}_{2r}(z)
\end{cases}.
\label{eq17}
\end{equation}
After the insertion of the expression of $|S_{fi}|^{2}$, the decay rate $ d\Gamma $ becomes:
\begin{equation}
d\Gamma=\dfrac{G_{F}^{2}}{32Q_{1}}\dfrac{1}{(2\pi)^{5}}\sum_{r=-\infty}^{+\infty}\overline{|M_{fi}^{r}|}^{2}\int\dfrac{d^{3}q_{3}}{Q_{3}}\int\dfrac{d^{3}p_{2}}{E_{2}}\int\dfrac{d^{3}p_{4}}{E_{4}}\delta^{4}(p_{2}+p_{4}+q_{3}-q_{1}-rk),
\label{eq18}
\end{equation}
where
\begin{equation}
\overline{|M_{fi}^{r}|}^{2}=Tr\Big[(\slashed p_{2}+m_{N})\Delta^{r}(\slashed p_{1}+m_{P})\bar{\Delta^{r}}\Big] Tr\Big[(\slashed p_{4}+m_{\nu})\Lambda^{r}(\slashed p_{3}-m_{e})\bar{\Lambda^{r}}\Big].
\label{eq19}
\end{equation}
We begin the analytic calculation of the decay rate by integrating over $ d^{3}p_{4} $. We get:
\begin{equation}
d\Gamma=\dfrac{G_{F}^{2}}{32Q_{1}}\dfrac{1}{(2\pi)^{5}}\sum_{r=-\infty}^{+\infty}\overline{|M_{fi}^{r}|}^{2}\int\dfrac{d^{3}q_{3}}{Q_{3}}\int\dfrac{d^{3}p_{2}}{E_{2}}\dfrac{1}{E_{4}}\delta^{4}(E_{2}+E_{4}+Q_{3}-Q_{1}-r\omega),
\label{eq20}
\end{equation}
with $ E_{4}=\overrightarrow{|p_{4}|}=|r\omega+\overrightarrow{q_{1}}-\overrightarrow{p_{2}}-\overrightarrow{q_{3}}| $. In the proton's rest frame, we furthermore have $ Q_{1}=m_{p}^{*} $ and $ \overrightarrow{q_{1}}=0 $, and this leads to $ E_{4}=\overrightarrow{|p_{4}|}=|r\omega-\overrightarrow{p_{2}}-\overrightarrow{q_{3}}| $. In equation (\ref{eq20}), we replace $ d^{3}p_{2}$ by the following expression $ d^{3}p_{2}=|\mathbf{p_{2}}|E_{2}dE_{2}d\varphi_{2}d\cos(\theta_{2}) $. We obtain:
\begin{equation}
d\Gamma=\dfrac{G_{F}^{2}}{32Q_{1}}\dfrac{1}{(2\pi)^{5}}\sum_{r=-\infty}^{+\infty}\overline{|M_{fi}^{r}|}^{2}\int\dfrac{d^{3}q_{3}}{Q_{3}}\int_{0}^{2\pi}d\varphi_{2}|\mathbf{p_{2}}|dE_{2}\int_{-1}^{1}\delta^{4}(E_{2}+E_{4}+Q_{3}-Q_{1}-r\omega)\dfrac{d\cos(\theta_{2})}{E_{4}}.
\label{eq21}
\end{equation}
Let us put $ x=\cos(\theta_{2}) $ and denote the argument of the $ \delta $ function by $ g(x) $ which can be developed as follows:
\begin{eqnarray}
g(x)&=&\nonumber E_{2}+\Big[\omega^{4}+(E_{2}^{2}-m_{N}^{2})+(Q_{3}^{2}-m_{e}^{2})-2\omega^{2}\sqrt{E_{2}^{2}-m_{N}^{2}}x-2\omega^{2}\sqrt{Q_{3}^{2}-m_{e}^{2}}\cos(\theta_{1})\\&+& 2\sqrt{E_{2}^{2}-m_{N}^{2}}\sqrt{Q_{3}^{2}-m_{e}^{2}}\big[\cos(\varphi_{1})\sin(\theta_{1})\cos(\varphi_{2})\sqrt{1-x^{2}}\\&+&\nonumber\sin(\varphi_{1})\sin(\theta_{1})\sin(\varphi_{2})\sqrt{1-x^{2}}+\cos(\theta_{1})x\big]\Big]^{1/2}+Q_{2}-Q_{1}-r\omega .
\label{eq22}
\end{eqnarray}
We proceed by using the following standard formula:
\begin{equation}
\delta(g(x))=|g^{\prime}(x_{0})|^{-1}\delta(x-x_{0}),
\label{eq23}
\end{equation}
where $ x_{0} $ is the solution of the equation $ g(x)=0 $ and $ g^{\prime}(x_{0}) $ is the derivative of $ g(x) $ at $ x=x_{0} $, and it is expressed as:
\begin{eqnarray}
g^{\prime}(x_{0})&=&\nonumber -\dfrac{1}{E_{4}}\Big[\omega^{2}\sqrt{E_{2}^{2}-m_{N}^{2}}+\sqrt{E_{2}^{2}-m_{N}^{2}}\sqrt{Q_{3}^{2}-m_{e}^{2}}\big[\cos(\varphi_{1})\sin(\theta_{1})\cos(\varphi_{2})\Big(\dfrac{x}{\sqrt{1-x^{2}}}\Big)\\&+& \sin(\varphi_{1})\sin(\theta_{1})\sin(\varphi_{2})\Big(\dfrac{x}{\sqrt{1-x^{2}}}\Big)+\cos(\theta_{1})\big]\Big],
\label{eq24}
\end{eqnarray}
which gives us immediately:
\begin{equation}
d\Gamma=\dfrac{G_{F}^{2}}{32Q_{1}}\dfrac{1}{(2\pi)^{5}}\int\dfrac{d^{3}q_{3}}{Q_{3}}\int_{\frac{Q_{1}}{2}-Q_{3}}^{\frac{Q_{1}}{2}}|\mathbf{p_{2}}|dE_{2}\int_{0}^{2\pi}d\varphi_{2}\dfrac{1}{|g^{\prime}(x)|_{g(x)=0}}\sum_{r=-\infty}^{+\infty}\overline{|M_{fi}^{r}|}^{2}.
\label{eq25}
\end{equation}
In the last expression of $d\Gamma$, we use $ d^{3}q_{3}=|\mathbf{q_{3}}|Q_{3}dQ_{3}d\varphi_{1}d\cos(\theta_{1}) $. Now, we can get the total decay rate by performing the final integration over $ Q_{3} $ from zero to $ Q_{1}/2 $, hence:
\begin{equation}
\Gamma=\dfrac{G_{F}^{2}}{32Q_{1}}\dfrac{1}{(2\pi)^{5}}\int_{0}^{\frac{Q_{1}}{2}}|\mathbf{q_{3}}|dQ_{3}\int_{-1}^{1}d\cos(\theta_{1})\int_{0}^{2\pi}d\varphi_{1}\int_{\frac{Q_{1}}{2}-Q_{3}}^{\frac{Q_{1}}{2}}|\mathbf{p_{2}}|dE_{2}\int_{0}^{2\pi}d\varphi_{2}\dfrac{1}{|g^{\prime}(x)|_{g(x)=0}}\sum_{r=-\infty}^{+\infty}\overline{|M_{fi}^{r}|}^{2}.
\label{eq26}
\end{equation}
The quantity $ Q_{1}=m_{P}^{*}=\sqrt{m_{p}^{2}+e^{2}a^{2}} $ represents the effective mass
of the charged proton acquired inside the electromagnetic field. The trace calculations are performed by using FEYNCALC program \cite{FeynCalc}.
The lifetime of the proton is defined by the following expression:
\begin{equation}
\tau_{P}=\dfrac{1}{\Gamma},
\label{eq27}
\end{equation}
where $ \Gamma $ is the total decay rate of the charged proton inside the electromagnetic field.
\section{RESULTS AND DISCUSSION}
In this section, we discuss and analyze the numerical results obtained about the decay process of the proton into neutron, positron and electron neutrino. The proton is considered in its rest frame where its mass is taken as $m_{p}= 938.27\times 10^{-3} GeV$, and the mass of the positron is  $m_{e}=0.511\,MeV$ \cite{PDG}. The laser wave is circularly polarized, and its direction is chosen along the $z$-axis. The evolution of the proton decay rate and its lifetime will be discussed as a function of the external field parameters such as its strength and frequency. We mention that in the following discussion, the decay rate and the lifetime are summed over a number of exchanged photon $r$ from $-40$ to $+40$.
Different known powerful laser sources are used such as the \textbf{CO$_{2}$} laser, \textbf{Nd:YAG} laser and the \textbf{He:Ne} laser, and which correspond to  the frequencies $0.117 eV$, $1.17 eV$ and $2 eV$, respectively.
To obtain the lifetime in second ($s$) and decay rate in $eV$, we apply the following formula:
\begin{equation}
\tau_{P}[s]=\dfrac{6.58212\times10^{-16}[eV.s]}{\Gamma[eV]}.
\label{eq28}
\end{equation}
It is well known that the decay process of the proton into neutron, positron and electron neutrino is forbidden in vacuum due to the fact that the proton's mass is less than that of the neutron. However, inside a laser field, the proton can acquire a supply of (mass) energy. Consequently, the process may be possible in the presence of strong laser pulse.
The table below shows the effect of the laser field of circular polarization on the mass that the proton can acquire inside this field (effective mass). 
\begin{table}[H]
 \centering
\caption{\label{tab1}Effective mass of the proton ($ m_{P}^{*}$) as a function of the laser field strength for different laser sources}
\begin{tabular}{cccc}
\hline
\hline
  &    &   $ m_{P}^{*}[GeV] $ & \\
  
\cmidrule(lr){2-4} $ \varepsilon_{0}[V.cm^{-1}] $ & $\mathbf{CO_{2}}$ Laser$ (\omega=0.117 eV) $   &   \textbf{Nd:YAG} Laser $ (\omega=1.17 eV) $ & \textbf{He:Ne} Laser $ ( \omega=2 eV ) $  \\
 \hline 
 $ 10 $ & $ 938.27\times 10^{-3} $ & $ 938.27\times 10^{-3} $& $ 938.27\times 10^{-3} $\\
   \vdots &          \vdots & \vdots   & \vdots\\
 $ 10^{10} $ & $ 938.27\times 10^{-3} $ & $ 938.27\times 10^{-3} $& $ 938.27\times 10^{-3} $\\
 $ 10^{11} $ & $ 938.42\times 10^{-3} $ & $ 938.27\times 10^{-3} $& $ 938.27\times 10^{-3} $\\
 $ 10^{12} $ & $ 953.30\times 10^{-3} $ & $ 938.42\times 10^{-3} $& $ 938.32\times 10^{-3} $\\
 $ 10^{13} $ & $ 192.99\times 10^{-2} $ & $ 953.308\times 10^{-3} $& $ 943.443\times 10^{-3} $\\
 $ 10^{14} $ & $ 168.91\times 10^{-1} $ & $ 192.998\times 10^{-2} $& $ 136.154\times 10^{-2} $\\
 $ 10^{15} $ & $ 168.66 $ & $ 168.92\times 10^{-1} $& $ 991.087\times 10^{-2} $\\
      \hline\hline
\end{tabular}
\end{table}
Different laser sources, namely the \textbf{Nd:YAG} laser, the \textbf{He:Ne} laser and the $\mathbf{CO_{2}}$ laser, are studied with different laser field strengths. According to this table, the proton doesn't acquire any effective mass for laser strengths less than the thresholds $ \varepsilon_{0} =10^{11}\,V.cm^{-1}$ , $ \varepsilon_{0} =10^{12}\,V.cm^{-1}$  and $ \varepsilon_{0} =10^{12}\,V.cm^{-1}$ for $\mathbf{CO_{2}}$ laser, the \textbf{Nd:YAG} laser and the for \textbf{He:Ne} laser, respectively. 
In addition, from these thresholds the effective mass of the proton begins to increase rapidly until it reaches $ m_{P}^{*}=168.66\,GeV $ for the $\mathbf{CO_{2}}$ laser at $ \varepsilon_{0} =10^{15}\,V.cm^{-1}$, and this effective mass is $180$ times greater than its rest mass. We also notice that the effective mass of the proton at $ \varepsilon_{0} =10^{15}\,V.cm^{-1}$ is $168.92\times 10^{-1}\,GeV$ and $991,087\times 10^{-2}\,GeV$ for the the \textbf{Nd:YAG} laser and for the \textbf{He:Ne} laser, respectively.
By a simple calculation, we find that the mass deficit which makes this process impossible to occur in vacuum is $m_{n}+m_{e}-m_{p}=1.8 MeV$, where $m_{n}$, $m_{e}$ and $m_{p}$ are the masses of neutron, positron and proton. This deficit can be re-compensated by applying a $\mathbf{CO_{2}}$ laser with a strength greater or equal to $\varepsilon_{0} =10^{13}\,V.cm^{-1}$. The Schwinger limit for a laser field is given by $ \varepsilon_{limit}=m_{e}^{3}c^{3}/e\hbar \approx 1.3\times 10^{16}\,V.cm^{-1}$, and at this limit the electron–positron pairs start to be produced from vacuum.
Therefore,  the range of the laser wave strengths at which the results obtained have a physical meaning is approximately between $10^{13}\,V.cm^{-1}$ and $9\times 10^{15}\,V.cm^{-1}$.
\begin{figure}[H]
  \centering
      \includegraphics[scale=0.58]{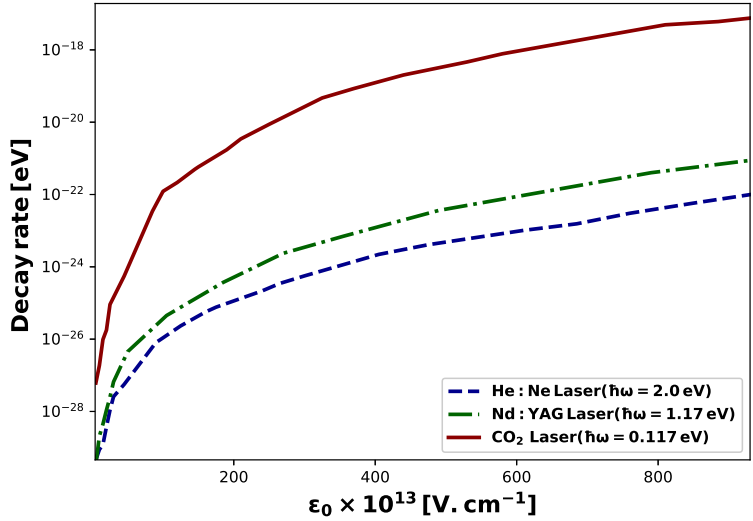}
        \caption{Variation of the proton's decay rate as a function of the laser field strength for different laser frequencies.}
        \label{fig1}
\end{figure}
Obviously, since the proton acquires an effective mass inside the electromagnetic field, then its decay into neutron, positron and electron neutrino may become possible. Besides, its lifetime and decay rate change with respect to the laser field strength. In figure \ref{fig1}, we show the variation of the proton's decay rate as a function of the wave strength of different laser sources.
As discussed above, the laser field strength is chosen to be ranging from $10^{13}\,V.cm^{-1}$ and $9\times 10^{15}\,V.cm^{-1}$. It is obvious from the figure \ref{fig1} that for all laser sources the decay rate of the proton decay into into neutron, positron and electron neutrino increases progressively by increasing the laser strength.
For instance, for $\omega=2 \,eV$, the decay rate is $ \Gamma =4.583\times 10^{-30}\,eV$ and $\Gamma =4.932\times 10^{-24}\,eV$ successively at $\varepsilon_{0}=5\times 10^{13}\,V.cm^{-1}$ and $\varepsilon_{0}=500\times 10^{13}\,V.cm^{-1}$.
Moreover, for the same laser strength, the order of magnitude of the decay rate differs from one frequency to another, and it increases as much as the laser source frequency decreases.
For example, at $\varepsilon_{0}=500\times 10^{13}\,V.cm^{-1}$, the decay rate is $ \Gamma =3.250\times 10^{-19}\,eV$, $ \Gamma =3.785\times 10^{-23}\,eV$ and $ \Gamma =4.932\times 10^{-24}\,eV$ for the $\mathbf{CO_{2}}$ laser, \textbf{Nd:YAG} laser  and the \textbf{He:Ne} laser, respectively.
As a result, the $\mathbf{CO_{2}}$ laser seems to have a great impact on the decay rate from low laser strengths as compared to the other laser sources.
\begin{figure}[H]
  \centering
      \includegraphics[scale=0.58]{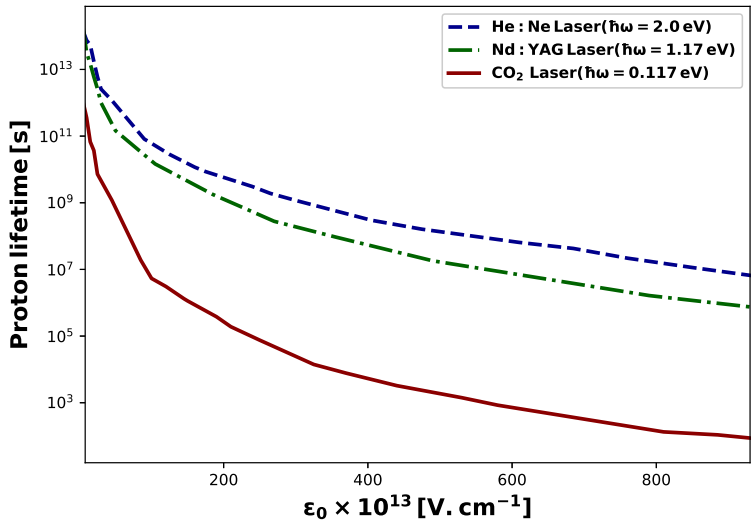}
        \caption{Lifetime of the proton versus the laser field strength for different laser sources.}
        \label{fig2}
\end{figure}
Figure \ref{fig2} illustrates the dependence of the proton's lifetime on the electric field strength $\varepsilon_{0}$. Again the range of laser field strengths is taken between $10^{13}\,V.cm^{-1}$ and the Schwinger limit. As expected from equation (\ref{eq28}), the lifetime of the proton begins to decrease significantly from $ \varepsilon_{0} =10^{13}\,V.cm^{-1}$. 
For example, the values of the lifetime that correspond to the frequency $\omega=0.117\,eV$ are $879\,s$, $685\,s$ and $133\,s$ for $ \varepsilon_{0} = 580\times 10^{13}\,V.cm^{-1}$, $ \varepsilon_{0} = 600\times 10^{13}\,V.cm^{-1}$ and $ \varepsilon_{0} = 800\times 10^{13}\,V.cm^{-1}$, respectively.
In addition, this decreasing process of the lifetime as a function of the laser strength depends also on the frequency of the laser pulse. 
Indeed, low laser frequencies lead to a great decrease of the proton's lifetime. For instance, for $ \varepsilon_{0} = 400\times 10^{13}\,V.cm^{-1}$, the proton's lifetime is $4889.6\,s$, $4.7\times 10^{7}\,s$ and $2.62\times 10^{8}\,s$ for the frequencies $0.117\,eV$, $1.17\,eV$ and $2\,eV$, respectively. 
Therefore, we assume that the $\mathbf{CO_{2}}$ laser is the most convenient laser source in this case.
Consequently, by using the $\mathbf{CO_{2}}$ laser with circular polarization, the lifetime of the proton can be shortened significantly without overcoming the Schwinger limit. More precisely, with the $\mathbf{CO_{2}}$ laser, the lifetime of the proton is $879\,s$ at $ \varepsilon_{0} = 580\times 10^{13}\,V.cm^{-1}$ which is approximately equal to that of the neutron \cite{neutron}. In this respect, it is possible to turn the proton, which is an unstable particle in the standard model, into an unstable particle decaying into heavier particle such as neutron in association with a positron and electron neutrino.
\section{Conclusion}
The decay of the proton inside an external field into neutron, positron and electron neutrino is investigated in this paper. The lifetime of the proton and its probability to decay into neutron, positron and electron neutrino are calculated analytically by using Dirac-Volkov formalism and the S-matrix approach. Then, these physical quantities are numerically computed, analyzed and discussed for different laser sources and for different intensities. 
We have found that laser wave with circular polarization has a great effect on the lifetime and decay rate of the proton especially for laser field strengths between $ \varepsilon_{0} = 10^{13}\,V.cm^{-1}$ and $ \varepsilon_{0} = 9\times 10^{15}\,V.cm^{-1}$ (near the Schwinger limit). 
In addition, using low frequency laser pulse lead to more significant results. For instance, the $\mathbf{CO_{2}}$ laser with the strength $ \varepsilon_{0} = 580\times  10^{13}\,V.cm^{-1}$ decreases the lifetime of the proton until it becomes equal to that of the neutron ($\approx 879\,s$). With the current breakthroughs in laser technology, we hope that this result will be tested experimentally.

\end{document}